\documentclass[english,reprint,pra,aps,longbibliography,superscriptaddress]{revtex4-1}
\usepackage[T1]{fontenc}
\usepackage[utf8]{inputenc}
\usepackage{babel}
\usepackage{amsmath}
\usepackage{amsthm}
\usepackage{amssymb}
\usepackage{graphicx}
\usepackage{amsfonts}

\RequirePackage[dvipsnames,usenames]{xcolor}
\definecolor{pdarkblue}{rgb}{0.1797, 0.1875, 0.5703}

\usepackage[unicode=true,bookmarks=false,breaklinks=false,pdfborder={0 0 1},backref=false,colorlinks=true,citecolor=pdarkblue,linkcolor=pdarkblue,urlcolor=pdarkblue]{hyperref}

\makeatletter

\theoremstyle{plain}
\newtheorem*{thm*}{\protect\theoremname}\theoremstyle{plain}
\theoremstyle{plain}

\usepackage{newtxtext}

\DeclareMathOperator*{\argmax}{\arg\,\max}
\DeclareMathOperator{\atanh}{\tanh^{-1}}

\global\long\def\evndx{\alpha}%
\global\long\def\imag{\mathrm{Im}\,}%
\global\long\def\RR{W}%
\global\long\def\symmR{\bar{\RR}}%
\global\long\def\symmRij{\symmR_{ij}}%
\global\long\def\symmRji{\symmR_{ji}}%
 
\global\long\def\epr{\sigma}%
 
\global\long\def\eigvec{\boldsymbol{\lambda}}%
\global\long\def\rre{\mathrm{Re}\,}%

\global\long\def\eprR{\sigma}%
\global\long\def\eqBnd{\sqrt{\frac{\constZ}{2}\epr}}

\global\long\def\deltaEV{\Vert\Delta\eigvec\Vert}%
\global\long\def\ii{\mathrm{i}}%
\global\long\def\constZ{\kappa}%
\global\long\def\constZepr{\constZ\eprR}

\global\long\def\eigvec{\boldsymbol{\lambda}}%
\global\long\def\eigvecW#1{\boldsymbol{\lambda}^{#1}}%
\global\long\def\eigvecWi#1#2{\lambda_{#2}^{#1}}%
\global\long\def\relL{\vert\rre\lambda_{2}^{\RR}\vert}%

\global\long\def\constB{\eta_{\RR}}%
\global\long\def\constG{\eta_{G}}%

\global\long\def\TTt{T}
\global\long\def\TTtji{T_{ji}}
\global\long\def\TTtij{T_{ij}}

\global\long\def\ppvec{\boldsymbol{p}}
\global\long\def\pivec{\boldsymbol{\pi}}
\global\long\def\onevec{\boldsymbol{1}}
\global\long\def\phivec{\boldsymbol{\phi}}
\global\long\def\vvec{\boldsymbol{v}}
\global\long\def\uvec{\boldsymbol{u}}

\global\long\def\dyntau{\tau}

\global\long\def\act{\mathcal{A}}%
\global\long\def\actji{\act_{ji}}%

\newcommand\refboundfig{Fig.~\ref{fig:1}(b)}
\newcommand\refpushpullfig{Fig.~\ref{fig:1}(a)}

\makeatother

\begin{document}
\title{A thermodynamic bound on spectral perturbations,\\
with applications to oscillations and relaxation dynamics}
\author{Artemy Kolchinsky}
\email{artemyk@gmail.com}

\affiliation{ICREA-Complex Systems Lab, Universitat Pompeu Fabra, 08003 Barcelona, Spain
\looseness=-1
}
\affiliation{Universal Biology Institute, Graduate School of Science, The University of
Tokyo, 7-3-1 Hongo, Bunkyo-ku, Tokyo 113-0033, Japan
\looseness=-1
}
\author{Naruo Ohga}
\affiliation{Department of Physics, Graduate School of Science, The University
of Tokyo, 7-3-1 Hongo, Bunkyo-ku, Tokyo 113-0033, Japan
\looseness=-1
}
\author{Sosuke Ito}
\affiliation{Universal Biology Institute, Graduate School of Science, The University of
Tokyo, 7-3-1 Hongo, Bunkyo-ku, Tokyo 113-0033, Japan
\looseness=-1
}
\affiliation{Department of Physics, Graduate School of Science, The University
of Tokyo, 7-3-1 Hongo, Bunkyo-ku, Tokyo 113-0033, Japan
\looseness=-1
}
\begin{abstract}
In discrete-state Markovian systems, many important properties of correlations functions and relaxation dynamics depend on the spectrum of the rate matrix. 
Here we demonstrate the existence of a universal trade-off between thermodynamic and spectral properties. We show that the entropy production rate,
the fundamental thermodynamic cost of a nonequilibrium steady state, 
bounds the difference between the eigenvalues of a  nonequilibrium rate matrix
and a reference equilibrium rate matrix. Using this result, we
derive thermodynamic bounds on the spectral gap, which governs autocorrelations times and the speed of relaxation to steady state. We also derive thermodynamic bounds on the imaginary eigenvalues, which govern the speed of oscillations. We illustrate our approach using a simple model of biomolecular sensing. 
\end{abstract}
\maketitle

\section{Introduction}

One of the main goals of nonequilibrium thermodynamics is to understand
trade-offs between thermodynamic costs and functionality in molecular
systems~\cite{dou2019thermodynamic,mehta2016landauer}. 
The entropy production rate (EPR) is one of the most important costs, since it quantifies both thermodynamic dissipation~\cite{van2015ensemble} { and statistical irreversibility~\cite{fodor2022irreversibility}}. It has been found to constrain functional
properties such as accuracy~\cite{hopfield1974kinetic,sartori2015thermodynamics},
sensitivity of nonequilibrium response~\cite{owen2020universal},
and precision of fluctuating observables~\cite{horowitz2020thermodynamic}.

{
Here we consider discrete-state Markovian systems, which are commonly employed in stochastic thermodynamics, biophysics, chemistry, and other fields~\cite{schnakenbergNetworkTheoryMicroscopic1976,van1992stochastic,allen2010introduction}. In such systems, many important properties are determined by the spectrum (set of eigenvalues) of the rate matrix. 
For concreteness, one may imagine a molecular system that transitions between a finite number of configurations, such as the push-pull model shown in \refpushpullfig{} and studied in our examples below. In general, the system is described by a probability distribution $\ppvec(t)=(p_{1}(t),\dots,p_{n}(t))$ with dynamics $\partial_t \ppvec(t) = \RR \ppvec(t)$, where $\RR$ is the rate matrix. 
The system's evolution over time $\dyntau$ is given by $\ppvec(\dyntau) = e^{\dyntau \RR} \ppvec(0)$. Assuming $\RR$ is irreducible and diagonalizable~\footnote{If $\protect \RR$ is not diagonalizable, it can still be written using the Jordan-Chevalley decomposition as the sum of two commuting matrices $\protect \RR=\RR'+N$, with $\protect \RR'$ diagonalizable and $\protect N$ nilpotent. Then $\protect e^{\dyntau\RR}=e^{\dyntau\RR'}e^{\dyntau N}$, where $\protect e^{\dyntau\RR'}$ can be decomposed as in Eq.~(\ref{eq:transMx}) while $\protect e^{\dyntau N}$ contributes a factor that is polynomial in $\protect \dyntau$~\cite{goudon2017math}. A sufficient (but not necessary) condition for $\RR$ to be diagonalizable is having non-degenerate eigenvalues.}, the time-evolution operator can be expressed as \nocite{goudon2017math} 
\begin{align}
e^{\dyntau \RR}=\pivec \onevec^T+\sum_{\evndx: \lambda_\evndx \ne 0} e^{\dyntau\rre\lambda_{\evndx}}e^{\ii \dyntau\imag\lambda_{\evndx}} \boldsymbol{u}^{(\evndx)} {\boldsymbol{v}^{(\evndx)}}^T\,,
\label{eq:transMx}
\end{align}
where $\pivec$ is the steady-state distribution while $(\lambda_\evndx,\boldsymbol{u}^{(\evndx)},\boldsymbol{v}^{(\evndx)})$ refer to the eigenvalues and right/left eigenvectors of $\RR$.  
In this way, relaxation toward steady state is decomposed into contributions from different eigenmodes, with
mode $\evndx$ decaying with rate $-\rre\lambda_{\evndx} \ge 0$
and oscillatory frequency $\imag\lambda_{\evndx}/2\pi$~\cite{barato2017coherence,oberreiter2022universal,hodas2010quality}. 
The spectrum of the rate matrix also determines other aspects of relaxation. For instance, degenerate eigenvalues can lead to power-law decay times~\cite{polettiniFisherInformationMarkovian2014a,andrieuxSpectralSignatureNonequilibrium2011} as well as dynamical phase transitions in relaxation trajectories~\cite{teza2023eigenvalue}.

The spectrum also plays a role in autocorrelation functions. The steady-state autocorrelation of observable $a$ over time lag $\dyntau$ %
is 
\begin{align}
\langle a(\dyntau)a(0)\rangle-\langle a\rangle\langle a\rangle=\sum_{i,j}([e^{\dyntau \RR}]_{ji}-\pi_{j})\pi_{i}a_{j}a_{i}.
\label{eq:autocorr}
\end{align} 
Combining Eqs.~\eqref{eq:transMx} and~\eqref{eq:autocorr} %
implies that the real parts of the eigenvalues control the decay rates of autocorrelations, while the imaginary parts control their oscillations~\cite{qian2000pumped,ohga2023thermodynamic}.

Relaxation and autocorrelation timescales determine many important functions in biomolecular systems,
including the response speed of sensors to changing environmental parameters~\cite{mancini2013time}, the information-theoretic capacity of signal transmission~\cite{munakata2006stochastic,tostevin2009mutual}, and the robustness of bistable states~\cite{jia2020dynamical}. 
Especially important is the so-called \emph{spectral gap}, the smallest nonzero $-\rre\lambda_{\evndx}$, which determines the slowest timescale~\cite{mitrophanov2004spectral}. 
Oscillatory behavior also has important functional functions, e.g., in biochemical clocks~\cite{barato2017coherence,del2020high,uhlAffinitydependentBoundSpectrum2019,oberreiter2021stochastic,oberreiter2022universal,ohga2023thermodynamic,shiraishi2023entropy,zheng2023topological}. 

}

\begin{figure}[b]
\includegraphics[width=.9\columnwidth]{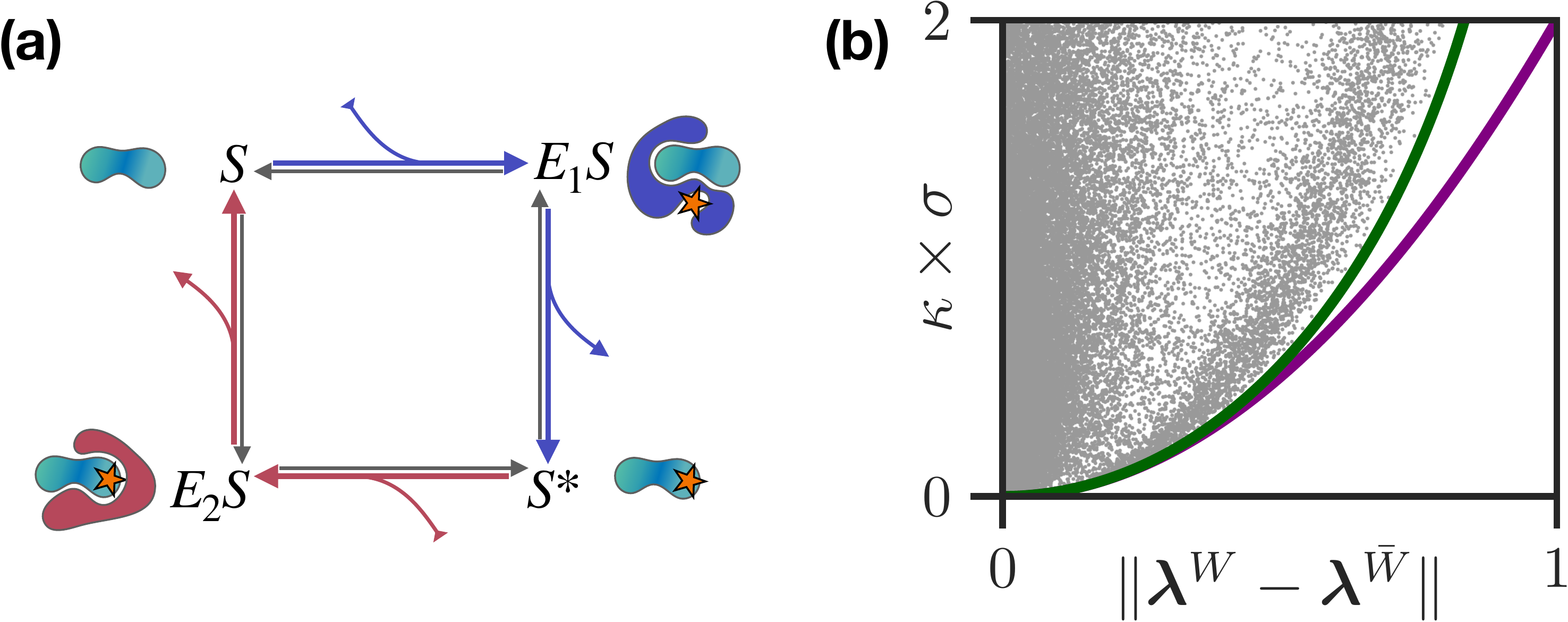}
\caption{\label{fig:1}
(a) {Our results imply thermodynamic trade-offs with relaxation speed and oscillations in discrete-state systems}, such as the 4-state ``push-pull'' model of enzyme concentration sensing~\cite{owen2020universal}.
(b) Our thermodynamic bounds on the spectral perturbation~\eqref{eq:boundFE} 
are illustrated on randomly sampled rate matrices. The far-from-equilibrium bound $\kappa \times \sigma \ge 2\deltaEV\atanh\deltaEV$
is shown in green and the near-equilibrium bound $\kappa \times \sigma \ge 2\deltaEV^{2}$ is shown in purple. 
}
\end{figure}

{
Given the importance of the spectrum in various functional properties, we ask whether there is any
relationship between a system's spectral and thermodynamic properties.  In this paper, we prove the existence of this kind of relationship and explore some of its physical consequences. 
As one consequence, we demonstrate that steady-state EPR bounds the size of the spectral gap, implying a fundamental thermodynamic cost for fast relaxation and for 
rapid decay of autocorrelation.
We also show that steady-state EPR bounds the size of the imaginary part of the spectrum, implying a fundamental thermodynamic cost for fast oscillations. Both consequences follow from a general relationship between thermodynamic cost and spectral perturbations, illustrated in \refboundfig{}. 
This general relationship shows that the steady-state
EPR under the nonequilibrium rate matrix $\RR$ bounds the difference between the spectrum
of $\RR$ and a ``reference'' equilibrium rate matrix $\symmR$, which has the same
steady state and dynamical activity as $\RR$ but obeys
detailed balance.}

\section{Preliminaries}
Given a rate matrix $\RR$ with steady-state $\pivec$, the steady-state EPR %
is defined as~\cite{schnakenbergNetworkTheoryMicroscopic1976,van2015ensemble}
\begin{equation}
\eprR:=\frac{1}{2}\sum_{i\ne j}(\pi_{i}\RR_{ji}-\pi_{j}\RR_{ij})\ln\frac{\pi_{i}\RR_{ji}}{\pi_{j}\RR_{ij}}\ge0\,.\label{eq:epr}
\end{equation}
For simplicity, we assume that $\RR$ is irreducible and  ``weakly reversible'' ($\RR_{ji}>0$ whenever $\RR_{ij}>0$) ensuring that $\pivec$ is unique and $\eprR$ is finite. 
{The steady-state EPR vanishes when forward and backward probability fluxes balance for all $i,j$, $\pi_{i}\RR_{ji}=\pi_{j}\RR_{ij}$. This condition is termed detailed balance (DB), and we identify it with equilibrium.}

\footnotetext[100]{Our spectral perturbation bounds are applicable to coarse-grained systems that are Markovian, so that their relaxation dynamics and autocorrelation functions can still be characterized by a rate matrix.}

{The quantity $\eprR$ can be interpreted as the rate of increase of thermodynamic entropy when the master equation obeys ``local detailed balance'' (LDB). Formally, LDB holds if for each transition $i\to j$, the log-ratio $\ln({\RR_{ji}}/{\RR_{ij}})$ %
can be identified with the corresponding increase of thermodynamic entropy in the environment. 
(See Ref.~\cite{maes_ldb_notes} for more details and a microscopic derivation of LDB.)  
LDB can also be considered for coarse-grained systems~\cite{Note100,esposito2012stochastic} and systems in contact with multiple thermodynamic reservoirs~\cite{esposito2010three}, in which case $\eprR$ provides a lower bound on the rate of increase
of thermodynamic entropy.
If LDB does not hold, $\eprR$ can still serve as a useful information-theoretic measure of
time irreversibility~\cite{fodor2022irreversibility}.}

In our first result below, we relate $\eprR$ to  %
the difference between the spectrum of the rate
matrix $\RR$ and that of a ``reference'' rate matrix $\symmR_{ij} :=(\RR_{ij}+\RR_{ji} {\pi_{i}}/{\pi_{j}})/2$. 
The rate matrix $\symmR$ obeys DB while having many of the same dynamical properties as $\RR$: 
the same steady state $\pivec$, same escape rates $\symmR_{ii}=\RR_{ii}$, 
and same \emph{dynamical activity}  across each transition. The dynamical activity across transition $i\to j$ is the rate of back-and-forth jumps~\cite{maes2020frenesy}, %defined as
{
\begin{align}
    \actji :=\pi_{i}\RR_{ji}+\pi_{j}\RR_{ij}=\pi_{i}\symmRji+\pi_{j}\symmRij.
    \label{eq:actdef}
\end{align}}
The rate matrix $\symmR$ is the equilibrium analogue of $\RR$, and it is equal to $\RR$ if and only if $\RR$ obeys DB. It is sometimes called the ``additive reversibilization'' of $\RR$ in the literature~\cite{sakai2016eigenvalue,bremaud2001markov}. 

{
For any rate matrix such as $\RR$, we use $\eigvecW \RR$ to indicate the vector of eigenvalues
of $\RR$ sorted in descending order by real part, $\rre\eigvecWi \RR 1\ge\dots\ge\rre\eigvecWi \RR n$. 
We will study the spectral perturbation between $\RR$ and $\symmR$, i.e., the difference between their eigenvalues. 
There are some subtleties involved in quantifying this difference, 
since in general there is no
canonical one-to-one mapping between the eigenmodes of these two matrices. Following previous work~\cite{kahanSpectraNearlyHermitian1975,bhatiaMatrixAnalysis1997}, we quantify it in terms of the distance between the sorted eigenvalue vectors:
$\deltaEV:=\Vert\eigvecW{\RR}-\eigvecW{\symmR}\Vert$. 
Importantly, $\deltaEV$ 
is the minimal distance across all possible one-to-one mappings between the eigenvalues,  
$\deltaEV=\min_\Pi \Vert\eigvecW{\RR}-\Pi\eigvecW{\symmR}\Vert$, where $\Pi$ runs over all permutation matrices~\footnote{This follows because $\protect \symmR$ has a real-valued spectrum. For details, see \cite[p.~166]{bhatiaMatrixAnalysis1997}.}. In physical terms, $\deltaEV$ compares the slowest modes of $\RR$ with the slowest modes of $\symmR$, and the fastest modes with the fastest modes.}

Finally, we introduce two rate constants that will play a central role in our analysis. 
The first rate constant, which we term the \emph{mixing
rate}, is defined
as 
\begin{align}
\constZ & := \max_{i\ne j}\frac{\actji}{2\pi_{i}\pi_{j}}.
\label{eq:cdef}
\end{align}
The mixing rate quantifies the maximum dynamical activity across any edge, after an appropriate normalization by the steady-state probabilities.  
As shown in Appendix~\ref{app:mixingrate}, it can be understood as the maximum speed with which probability
can flow between different regions of state space.  The second rate constant is
\begin{equation}
\constB:=\sqrt{\sum_{i\ne j}\pi_{i}\pi_{j}\Big(\frac{\actji}{2\pi_{i}\pi_{j}}\Big)^{2}}.\label{eq:normdef}
\end{equation}
It reflects the overall magnitude of the normalized dynamical
activity, where the contribution of each transition is weighted
by the steady-state distribution.
These two constants
depend only on the dynamical activity and steady-state distribution
and can be equivalently defined either
in terms of $\RR$ or $\symmR$.

\section{Thermodynamic bound on spectral perturbations}

In our first main result, we demonstrate that %
the steady-state EPR bounds 
the magnitude of the spectral perturbation $\deltaEV=\Vert\eigvecW{\RR}-\eigvecW{\symmR}\Vert$ as
\begin{align}
\frac{\eprR}{2\constZ} & \ge \left( \frac{\constB}{\constZ}\right)^2\frac{\deltaEV}{\constB}\atanh\frac{\deltaEV}{\constB}\ge \frac{\deltaEV^2}{\constZ^2} \,.\label{eq:boundFE}
\end{align}
{
The first bound is derived using a classic spectral perturbation theorem from linear algebra~\cite{kahanSpectraNearlyHermitian1975}, see derivation in Appendix~\ref{app:mainresult}.}
The second bound follows from $x \atanh x\ge x^2$, which is tight for small $x$. 
The two bounds
become equivalent near equilibrium, when $\eprR\approx 0$, $\RR\approx\symmR$, 
and $\deltaEV\approx0$.

Eq.~\eqref{eq:boundFE} implies that there is a universal thermodynamic cost for having the eigenvalues
of $\RR$ be different from those of its equilibrium analogue $\symmR$.  
{
Importantly, all terms that appear in our bounds ($\eprR$, $\constZ$, $\constB$, and $\deltaEV$) are functions of the system's rate matrix and they all have units of  ``per time''. Moreover, all of these terms increase in the same linear manner when the rate matrix is scaled as $\RR\mapsto \alpha \RR$. Therefore, the bounds~\eqref{eq:boundFE} are invariant to the overall scaling of the rate matrix, depending instead only on its structure. 

Observe that steady-state EPR~\eqref{eq:epr} depends both on the system's kinetic properties, via the fluxes $\pi_{i}\RR_{ji}-\pi_{j}\RR_{ij}$, and its thermodynamic properties, via the dimensionless ratios $\ln(\pi_{i}\RR_{ji}/\pi_{j}\RR_{ij})$ (which are called ``thermodynamic forces'' in the literature~\cite{esposito2010three}). 
The rate constants $\constB$ and $\constZ$ serve as normalization terms for the kinetics, allowing EPR to be compared to the spectral perturbation in a dimensionless manner. Near equilibrium, only the mixing rate $\constZ$ plays a role, while the rate constant $\constB$ becomes relevant in the far-from-equilibrium regime. The dimensionless ratio ${\constB}/{\constZ}$ in~\eqref{eq:boundFE} measures the disorder of the normalized dynamical activity, and it reaches its maximum when the values of $\actji/{2\pi_{i}\pi_{j}}$ are equal for all allowed transitions.}

We illustrate our result in \refboundfig{}. We generate
$10^{5}$ random 5-by-5 rate matrices, scaled to have $\constB=1$.
We plot $\constZ\times\eprR$ versus two lower bounds that follow from \eqref{eq:boundFE}: $2\deltaEV\atanh\deltaEV$
and $2\deltaEV^{2}$. Observe that the first
bound can be arbitrarily tight far from equilibrium. In fact, one can verify that it is saturated by unicyclic rate matrices with uniform rates.

We can also invert Eq.~\eqref{eq:boundFE} to 
derive an  upper 
bound on the spectral perturbation as a function of the EPR. Define the function $\Phi_{y} : \mathbb{R}_{\ge 0} \to \mathbb{R}_{\ge 0}$ as
\begin{equation}
\Phi_{y}(x):=2xy\atanh({x}/{y})\,,\label{eq:phixy}
\end{equation}
so that~\eqref{eq:boundFE} implies 
$\constZ\eprR\ge\Phi_{\constB}(\deltaEV)$.
We can then bound the spectral perturbation using EPR as
\begin{equation}
\deltaEV\le\Phi_{\constB}^{-1}(\constZepr)\le\eqBnd \,,\label{eq:invBounds}
\end{equation}
where $\Phi_{\constB}^{-1}$ is the inverse function (which can be
computed numerically from $\Phi_{\constB}$). 

\subsection{Intermediate bounds} 
The constant $\constB$ depends on the overall pattern of dynamical
activity across all transitions, which can be difficult to access in practice.  %
Below, it will be useful to derive more general bounds that do not depend on this constant. %
We define another rate
matrix $G$,
\begin{equation}
G_{ij}:=\begin{cases}
\pi_{i} & i\ne j,\,\RR_{ij}>0\\
0 & i\ne j,\,\RR_{ij}=0
\end{cases}\qquad G_{ii}:=-\sum_{j(\ne i)}G_{ji}.\label{eq:gdef}
\end{equation}
This rate matrix obeys DB and has the same steady-state distribution
and the same \emph{graph topology} as $\RR$ (the same pattern of allowed
transitions with non-zero rates). At the same time, it does not depend on the dynamical activity under $\RR$. The constant $\constB$ can then be bound as
\begin{equation}
\constB\le\kappa\constG\le\kappa\sqrt{1-\sum_{i}\pi_{i}^{2}}\le\kappa,\label{eq:seq}
\end{equation}
where $\constG$ is defined as in \eqref{eq:normdef} but using $G$ instead of $\RR$ (see Appendix~\ref{app:seq}).
Plugging these inequalities into \eqref{eq:invBounds} and
using that $\Phi_{y}^{-1}(x)$ is increasing in $y$ gives a hierarchy of inequalities 
which fall in-between the two expressions in
\eqref{eq:invBounds}:  %
\[
\deltaEV\le\Phi_{\constB}^{-1}(\constZepr)\le\Phi_{\kappa\constG}^{-1}(\constZepr)\le..\le\Phi_{\constZ}^{-1}(\constZepr)\le\eqBnd.
\]
Within this hierarchy, tighter bounds reflect finer-grained information
about the rate matrix $\RR$. The weakest bounds depend only on the mixing
rate $\constZ$, the intermediate
bound also depends on the steady-state distribution and the graph topology (via $\constG$), while the
tightest bound depends on the overall pattern of dynamical activity (via $\constB$).
In principle, there are systems in which all the bounds expressed in terms of $\Phi_{y}^{-1}$
are tight far-from-equilibrium.

\subsection{Real and imaginary eigenvalues}
The spectral perturbation can be decomposed into contributions from
the real and imaginary parts of each eigenvalue, 
\begin{align}
\deltaEV^{2}=\sum_{\evndx=2}^{n}\vert\Delta\rre\lambda_{\evndx}\vert^{2} +\sum_{\evndx=2}^{n}\vert\imag\eigvecWi{\RR}{\evndx}\vert^{2},\label{eq:evdecomp}
\end{align}
where %
$\Delta\rre\lambda_{\evndx}=\rre\eigvecWi{\RR}{\evndx}-\eigvecWi{\symmR}{\evndx}$.
In this expression, we used $\eigvecWi{\symmR}{1}=\eigvecWi{\RR}1=0$
and that the eigenvalues of $\symmR$ are
real valued (Appendix~\ref{app:spectral}). 

In the following, we use our general results to study separate trade-offs between EPR versus decay timescales and EPR versus oscillations. 
Interestingly, \eqref{eq:evdecomp} also implies a three-way trade-off between EPR, decay timescales, and oscillations, exploration of which we leave for future work.

\section{Spectral gap}
We now derive thermodynamic bounds on the spectral gap $\relL$, which controls the slowest decay timescale of relaxation and autocorrelation. %
 
We first consider a thermodynamic bound on the increase of the spectral gap:
\begin{equation}
0\le\relL-\vert\eigvecWi{\symmR}2\vert\le\Phi_{\constB}^{-1}(\constZepr)\le\eqBnd .\label{eq:acc}
\end{equation}
Thus, accelerating decay timescales under $\RR$, relative to the equilibrium rate
matrix $\symmR$, carries an unavoidable thermodynamic cost.  
Weaker but more general bounds can be derived by combining~\eqref{eq:acc} with
the inequalities \eqref{eq:seq}.

We derive~\eqref{eq:acc} by combining \eqref{eq:invBounds} with $\deltaEV\ge\vert\Delta\rre\lambda_{2}\vert$, then using 
that %
the spectral gap of $\RR$ is always larger
than the spectral gap of $\symmR$ (see Appendix~\ref{app:secev}):
\begin{align}
\vert\Delta\rre\lambda_{2}\vert=\relL-\vert\eigvecWi{\symmR}2\vert\ge0\,.\label{eq:fffd}
\end{align}
The inequality~\eqref{eq:fffd}
has been previously used to motivate ``irreversible Monte
Carlo'' schemes~\cite{ichikiViolationDetailedBalance2013,suwa2010markov,bierkens2016non,diaconis2000analysis,turitsyn2011irreversible,chen2013accelerating,takahashi2016conflict,sakai2016eigenvalue,kaiser2017acceleration,ghimenti2022accelerating}. 
We note that the upper bounds~\eqref{eq:acc} are often not tight, %
since the inequality $\deltaEV\ge\vert\Delta\rre\lambda_{2}\vert$
is usually not saturated.

It is often useful to bound the spectral gap of $\RR$ itself, rather
than the increase of the spectral gap of $\RR$ relative to $\symmR$.
To do so, we combine $\Phi_{\constB}^{-1}(\constZepr)\le\Phi_{\constZ\constG}^{-1}(\constZepr)$,
as follows from \eqref{eq:seq}, with the following inequalities
on the spectral gap of $\symmR$ (Appendix~\ref{app:ng2}):
\begin{equation}
\vert\eigvecWi{\symmR}2\vert\le\constZ\vert\eigvecWi G2\vert\le\constZ,\label{eq:ng2}
\end{equation}
where $\vert\eigvecWi G2\vert$ is the spectral gap of the rate matrix
$G$ from \eqref{eq:gdef}.  
Combining with \eqref{eq:acc}
and rearranging gives %
\begin{equation}
\relL\le\Phi_{\constZ\constG}^{-1}(\constZepr)+\constZ\vert\eigvecWi G2\vert\le\eqBnd +\constZ\vert\eigvecWi G2\vert.\label{eq:spectralgap-bound}
\end{equation}
These bounds depend only on the mixing rate $\constZ$, steady state $\pivec$, and graph topology of the rate matrix. 
They include one term that is a thermodynamic cost, which depends on the EPR and vanishes in
equilibrium, and a second ``baseline'' term, which does not vanish in equilibrium. Weaker but simpler bounds can also
be derived using \eqref{eq:seq} and \eqref{eq:ng2}. For instance,
we may derive bounds which depend only on the mixing
rate as
\begin{equation}
\relL\le\Phi_{\constZ}^{-1}(\constZepr)+\constZ\le\eqBnd +\constZ.\label{eq:sgap2}
\end{equation}
Eqs.~\eqref{eq:spectralgap-bound} and \eqref{eq:sgap2} together
form our second main result.

\subsection{Example}
We illustrate our bounds on the spectral gap  %
by studying the thermodynamic cost of response speed in biomolecular sensing~\cite{mehta2012energetic,govern2014energy,skoge2013chemical,owen2020universal,qian2003thermodynamic,goldbeter1981amplified}. 
We consider a 4-state system with a cyclic topology, \refpushpullfig{},
inspired by the ``push-pull'' enzymatic sensor in the regime of low substrate concentrations~\cite{owen2020universal}. Here, a substrate molecule can be either unmodified ${S}$ or modified ${S}^*$. Modification is catalyzed by enzyme $E_1$ via bound state $E_1 S$, and demodification is catalyzed by enzyme $E_2$ via bound state $E_2 S$. %

The steady-state distribution depends on the concentrations of $E_1$ and $E_2$ and serves as the sensor readout. 
Following a change of enzyme concentrations, the sensor's response speed is determined by the rate of relaxation toward the new steady state, which can be quantified using the spectral gap $\relL$. The same idea applies not only to the push-pull system but also other molecular sensors where the steady state reflects environmental parameters.

\begin{figure}
\includegraphics[width=1\columnwidth]{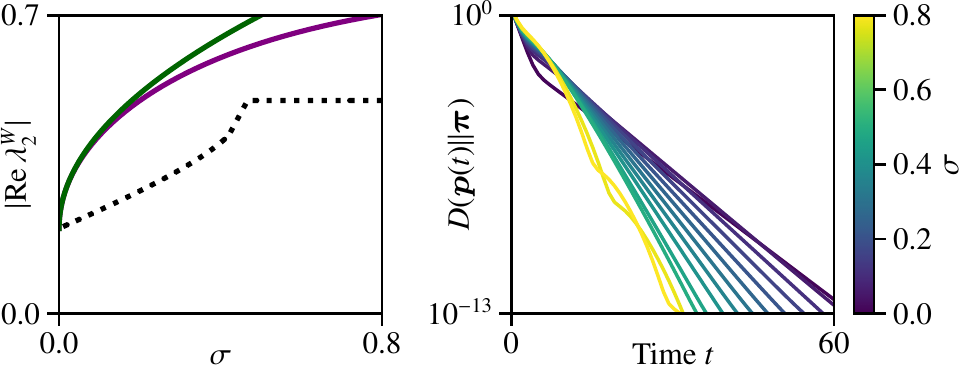} \caption{\label{fig:realpart}Thermodynamic bound on the
spectral gap illustrated on the biomolecular sensing
model from \refpushpullfig{}. \emph{Left}: Two bounds
from \eqref{eq:spectralgap-bound} (solid lines) versus the
maximal spectral gap (dotted line) identified by numerical optimization 
at varying levels of steady-state EPR. \emph{Right}: larger EPR is associated with faster relaxation
to steady state, shown using the decay of KL divergence $D(\ppvec(t)\Vert\pivec)$
over time.}
\end{figure}

We use numerical optimization to find
4-by-4 cyclic rate matrices that have the largest spectral gap for
a given steady-state EPR, given a fixed steady-state distribution $\pivec$ and %
mixing rate $\constZ$. For concreteness, we fix $\pivec=(0.4,0.1,0.4,0.1)$
and $\constZ=1$. 
Fig.~\ref{fig:realpart}(left) compares our bounds on the spectral gap~\eqref{eq:spectralgap-bound} with the largest gap found by numerical optimization at different levels of EPR. 
Fig.~\ref{fig:realpart}(right) also shows that increased steady-state
EPR is actually associated with faster relaxation. Specifically,
using the rate matrices identified using numerical optimization, we
plot $D(\ppvec(t)\Vert \pivec)=\sum_i p_i(t) \ln [p_i(t)/\pi_i]$. This is the Kullback--Leibler (KL) divergence between the time-dependent probability
distribution $\ppvec(t)$ (starting from $\ppvec(0)=(1,0,0,0)$) and the steady state $\pivec$. 

As seen in Fig.~\ref{fig:realpart}, our bounds are always satisfied but only saturate at $\epr = 0$. % (away from equilibrium, $\deltaEV\ge\vert\Delta\rre\lambda_{2}\vert$ is usually not tight). 
Interestingly, the spectral gap plateaus above $\eprR \approx 0.5$, when it reaches the largest value achievable for a given topology and steady state.  Beyond this point, additional EPR leads to changes in other eigenvalues and increases the imaginary part of $\lambda_2$, as evidenced by oscillatory relaxation dynamics that emerge at large EPR in Fig.~\ref{fig:realpart}(right).

In future work, it may be interesting to relate our thermodynamic bound on response speed to existing thermodynamic bounds on sensitivity (ability of small parameter changes to elicit large changes in the steady state)~\cite{mehta2012energetic,govern2014energy,skoge2013chemical,owen2020universal,qian2003thermodynamic}.

\section{Imaginary eigenvalues}
In our final set of main results, %
we 
show that the steady-state EPR bounds the magnitude of the entire
imaginary part of the spectrum of $\RR$,
\begin{align}
\Vert\imag\eigvecW{\RR}\Vert\le\Phi_{\constB}^{-1}(\constZepr)\le\eqBnd  ,\label{eq:bndim}
\end{align}
where $\Vert\imag\eigvecW{\RR}\Vert=\sqrt{\sum_{\evndx=2}^{n}\vert\imag\eigvecWi{\RR}{\evndx}\vert^{2}}$. 
This follows immediately from \eqref{eq:invBounds} and \eqref{eq:evdecomp}. 
Intermediate bounds which do not depend on the constant
$\constB$ %
can be derived using \eqref{eq:seq}. 

We can also bound the imaginary part of any particular eigenmode (e.g., the slowest
or the fastest). Consider the size of the imaginary part of any eigenvalue, $\vert\imag\eigvecWi{\RR}{\evndx}\vert$ for $\evndx\in\{2,\dots,n\}$.
Since nonreal eigenvalues come in conjugate pairs, there must be
another eigenmode $\evndx'$ with $\vert\imag\eigvecWi{\RR}{\evndx'}\vert=\vert\imag\eigvecWi{\RR}{\evndx}\vert$.
Given \eqref{eq:evdecomp}, we have $\Vert\imag\eigvecW{\RR}\Vert^{2}\ge2\vert\imag\eigvecWi{\RR}{\evndx}\vert^{2}$,
which is tight for $n\le4$ and becomes increasingly weak for systems with many states. Plugging into \eqref{eq:bndim} gives
\begin{align}
\vert\imag\eigvecWi{\RR}{\evndx}\vert\le\frac{1}{\sqrt{2}}\Phi_{\constB}^{-1}(\constZepr)\le\sqrt{\frac{\constZ}{4}\eprR}.\label{eq:imevbound3}
\end{align}

\begin{figure}
\includegraphics[width=1\columnwidth]{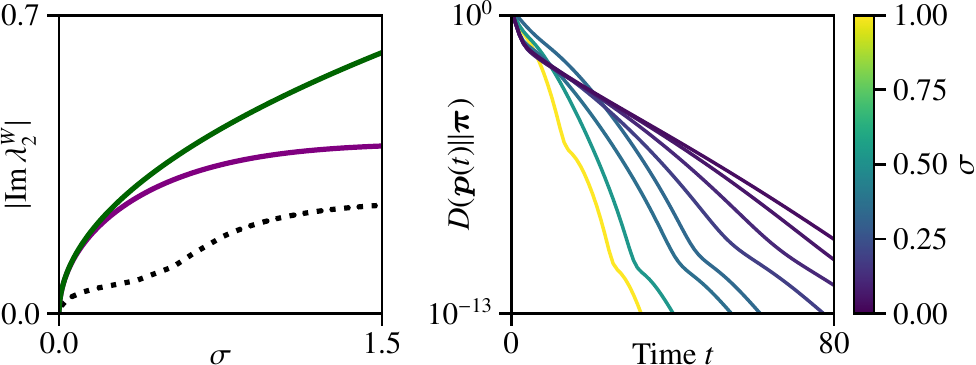} \caption{\label{fig:imag}Thermodynamic bound on imaginary eigenvalues illustrated on the model from \refpushpullfig{}. \emph{Left}: Two bounds
from \eqref{eq:imevbound3} (solid lines) versus the
maximal imaginary part of the second eigenvalue (dotted line) identified by numerical optimization 
at varying levels of steady-state EPR. \emph{Right}: larger EPR is associated with faster oscillatory ``pulsing'' during relaxation. 
}
\end{figure}

\subsection{Example} 
As an illustration, we study the trade-off between steady-state EPR and oscillatory behavior, as quantified by the imaginary part of the slowest eigenmode, $\vert\imag\eigvecWi{\RR}{2}\vert$. We again consider the 4-state cyclic system studied above, \refpushpullfig{}.  Although oscillatory behavior is not typically desired in biomolecular sensors, it is necessary for other types of systems modeled using similar cyclic rate matrices, such as biochemical clocks~\cite{barato2017coherence}.

We use numerical optimization to find
4-by-4 cyclic rate matrices that maximize $\vert\imag\eigvecWi{\RR}{2}\vert $ for
a given EPR, given a fixed steady-state distribution and %
mixing rate $\constZ$ (chosen as above). 
In Fig.~\ref{fig:imag}(left), we compare our bounds~\eqref{eq:imevbound3} with the largest imaginary part found by numerical optimization at different levels of steady-state EPR. 
In Fig.~\ref{fig:imag}(right), we select some rate matrices identified using numerical optimization and
plot $D(\ppvec(t)\Vert \pivec)=\sum_i p_i(t) \ln [p_i(t)/\pi_i]$. 
%the KL divergence between the relaxing probability
%distribution $\ppvec(t)$ and the steady state $\pivec$. 
Observe that relaxation dynamics exhibit oscillatory ``pulsing'', whose frequency increases with steady-state EPR.

\section{Discussion}
In this paper, we proved a relationship between the thermodynamic and
spectral properties of a rate matrix. Our results on imaginary eigenvalues contribute to the extensive literature on the relationship between dissipation and oscillations~\cite{barato2017coherence,del2020high,uhlAffinitydependentBoundSpectrum2019,oberreiter2021stochastic,oberreiter2022universal,ohga2023thermodynamic,shiraishi2023entropy,zheng2023topological}.
 
More surprising, perhaps, is our thermodynamic bound on the real eigenvalues, including the increase of the spectral gap. This is because EPR quantifies the violation of time-reversal 
symmetry, which at first glance appears to be unrelated to the real part
of the eigenvalues. These results contribute
to the growing interest in time-symmetric phenomena~\cite{maesTimeSymmetricFluctuationsNonequilibrium2006,maes2020frenesy,baiesi2015inflow}
and relaxation timescales~\cite{ghimenti2022accelerating,bao2023universal,teza2023eigenvalue,dechant2023thermodynamic,andrieuxSpectralSignatureNonequilibrium2011}
as signatures of nonequilibrium.

In future work, it may be interesting to consider nonlinear
systems, e.g., by studying the relationship between EPR and the Jacobian spectrum in deterministic chemical reaction networks. 
It is also interesting to extend our approach to continuous-state
systems (Fokker--Planck generators) and open quantum systems (Lindblad generators). 
Interestingly, a recent
preprint derived a different type of thermodynamic bound on relaxation in Fokker-Planck dynamics~\cite{dechant2023thermodynamic}; comparison with our approach is left for future work.

%\vspace{10pt}

\begin{acknowledgments} 
A.K.~acknowledges funding from 
European Union's Horizon 2020 research and innovation program under the Marie Sklodowska-Curie Grant Agreement No.~101068029. N.O.~is supported by JSPS KAKENHI Grant No. 23KJ0732. 
S.I.~is supported by JSPS KAKENHI Grants No. 19H05796, No. 21H01560, No. 22H01141, No. 23H00467, JST ERATO-FS Grant No. JPMJER2204, and UTEC-UTokyo FSI Research Grant Program. 
\end{acknowledgments}

\appendix

\renewcommand{\appendixname}{APPENDIX}
\newcommand{\appsection}[1]{\section{\MakeUppercase{#1}}}

\appsection{Meaning of the mixing rate $\constZ$}
\label{app:mixingrate}

\global\long\def\sA{\mathcal{X}}% 
\global\long\def\sB{\mathcal{Y}}% 

We show that the mixing rate $\constZ$, as defined in \eqref{eq:cdef}, can be understood as the maximum
speed of probability flow between different regions of state space. 

Consider any two
subsets of states $\sA\subseteq \{1,\dots,n\}$ and $\sB\subseteq \{1,\dots,n\}$. 
Let $\TTt=e^{t\RR}$ indicate the time-$t$ transition matrix, and let
$p_{\sA\to\sB}(t)=\sum_{i\in\sA,j\in\sB}\pi_{i}\TTtji$
indicate the probability that the steady-state process visits subset
$\sA$ at time $0$ and subset $\sB$ at time $t$. 
Since we assume that $\RR$ is irreducible, this probability
converges to $p_{\sA\to\sB}(\infty)=\sum_{i\in\sA,j\in\sB}\pi_{i}\pi_{j}$
in the long-time limit $t\to\infty$. Then, the mixing rate can be written as 
the normalized growth
rate of $p_{\sA\to\sB}(t)+p_{\sB\to\sA}(t)$, maximized across all
subsets:
\begin{equation}
\constZ=\max_{\sA,\sB}\frac{d}{dt}\frac{p_{\sA\to\sB}(t)+p_{\sB\to\sA}(t)}{p_{\sA\to\sB}(\infty)+p_{\sB\to\sA}(\infty)}\Big\vert_{t=0}.\label{eq:mixingrate0-1}
\end{equation}

\newcommand\localActij{\pi_{i}\RR_{ji}+\pi_{j}\RR_{ij}}

To show the equivalence of~\eqref{eq:cdef} and \eqref{eq:mixingrate0-1}, we first bound the numerator in \eqref{eq:mixingrate0-1}
as
\begin{align*}
 & \frac{d}{dt}[p_{\sA\to\sB}(t)+p_{\sB\to\sA}(t)]\big\vert_{t=0}\\
 & \qquad\quad =\frac{d}{dt}\left[\sum_{i\in\sA,j\in\sB}(\pi_{i}\TTtji+\pi_{j}\TTtij)\right]\Bigg\vert_{t=0}\\
 & \qquad\quad=\sum_{i\in\sA,j\in\sB}\localActij \\
 & \qquad\quad\le\left(\sum_{i\in\sA,j\in\sB}\pi_{i}\pi_{j}\right)\left(\max_{i\in\sA,j\in\sB}\frac{\localActij}{\pi_{i}\pi_{j}} \right)\\
 & \qquad\quad=\left(\sum_{i\in\sA,j\in\sB}\pi_{i}\pi_{j}\right)\left(\max_{i\in\sA \ne j\in\sB}\frac{\actji}{\pi_{i}\pi_{j}} \right).
\end{align*}
In the last line, we used the definition of dynamical activity~\eqref{eq:actdef}. We also restricted the maximization to $i\ne j$ by using the fact that $\act_{ii}\le 0$ (since $\RR_{ii} \le 0$).  
The denominator of the objective in \eqref{eq:mixingrate0-1} is given
by 
\[ 
p_{\sA\to\sB}(\infty)+p_{\sB\to\sA}(\infty)=2\sum_{i\in\sA,j\in\sB}\pi_{i}\pi_{j}.
\]
Combining gives the following upper bound on the right side of \eqref{eq:mixingrate0-1},
\[
\max_{\sA\ne \sB}\max_{i\in\sA\neq j\in\sB}\frac{\actji}{2\pi_{i}\pi_{j}} \le \max_{i\ne j}\frac{\actji}{2\pi_{i}\pi_{j}}.
\]
where the last expression is the definition of $\constZ$ from~\eqref{eq:cdef}. 
It can be verified that this upper bound is achieved by choosing $\sA=\{i^{*}\}$
and $\sB=\{j^{*}\}$ in~\eqref{eq:mixingrate0-1}, where 
\[
(i^{*},j^{*})\in\argmax_{i\ne j}\frac{\actji}{2\pi_{i}\pi_{j}}.
\]
Therefore, \eqref{eq:cdef} and \eqref{eq:mixingrate0-1} are equivalent.

\appsection{\MakeTextUppercase{Derivation of $\constB$ bounds}~(\ref{eq:seq})}
\label{app:seq}
 
We first derive the following bound:
\begin{align*}
\constB&=  \sqrt{\sum_{i\ne j}\pi_{i}\pi_{j}\Bigg(\frac{\pi_{i}\RR_{ji}+\pi_{j}\RR_{ij}}{2\pi_{i}\pi_{j}}\Bigg)^{2}}\\
&=  \sqrt{\sum_{i\ne j:\RR_{ji}>0}\pi_{i}\pi_{j}\Bigg(\frac{\pi_{i}\RR_{ji}+\pi_{j}\RR_{ij}}{2\pi_{i}\pi_{j}}\Bigg)^{2}}\\
 & \le\constZ\sqrt{\sum_{i\ne j:\RR_{ji}>0}\pi_{i}\pi_{j}}\,.
\end{align*}
In the second line, we used the assumption $\RR$ is weakly reversible ($\RR_{ij}>0$ whenever $\RR_{ji}>0$), and in the last line we used the definition of $\constZ$ from \eqref{eq:cdef}. 
Next, observe that 
\[
\sqrt{\sum_{i\ne j:\RR_{ji}>0}\pi_{i}\pi_{j}}=\sqrt{\sum_{i\ne j}\pi_{i}\pi_{j}\Bigg(\frac{\pi_{i}G_{ji}+\pi_{j}G_{ij}}{2\pi_{i}\pi_{j}}\Bigg)^{2}}=\constG,
\]
where we used the definition of the rate matrix $G$ \eqref{eq:gdef}.
Thus, we recover the first inequality in \eqref{eq:seq}. 

The second
and third inequalities follow from
\[
\sqrt{\sum_{i\ne j:\RR_{ji}>0}\pi_{i}\pi_{j}}\le\sqrt{\sum_{i\ne j}\pi_{i}\pi_{j}}=\sqrt{1-\sum_{i}\pi_{i}^{2}}\le1.
\]

\appsection{{Spectral inequalities}}
\label{app:spectral}

%\global\long\def\MM{C}% 
%\global\long\def\MMT{\MM^{T}}% 

\global\long\def\AA{L}% 
\global\long\def\BB{M}% 
\global\long\def\MM{K}% 

%\subsection{Preliminaries}

We begin by reviewing some of the notation and definitions introduced in the main text, along with a few useful results from linear algebra.

For any $n$-by-$n$ matrix $M$, we use the notation $\eigvec^{M}$ to indicate the vector of eigenvalues of matrix $M$ sorted in descending order by real part, $\rre\eigvecWi M 1\ge\dots\ge\rre\eigvecWi M n$.

Given the rate matrix $\RR$, we 
define the reference equilibrium matrix
\begin{align}
\symmR_{ij} &:=\frac{1}{2}\big(\RR_{ij}+\RR_{ji} \frac{\pi_{i}}{\pi_{j}}\big)\,.\label{eq:wbardef}
\end{align}
Assuming $\RR$ has a unique steady-state distribution $\pivec$, the first eigenvalue
is $\lambda_{1}^{\RR}=0$, and $\rre \lambda_\alpha^\RR < 0 $ for all $\alpha \in \{2,\dots,n\}$. The right eigenvector corresponding to $\lambda_1^\RR$ is $\pivec = (\pi_{1},\dots,\pi_{n})^{T}$
so that $\RR\pivec=0$. The corresponding left eigenvector is $\onevec=(1,\dots,1)^{T}$
so that $\onevec^{T}\RR=0$. Any other right eigenvectors are orthogonal
to $\onevec$, and any other left eigenvectors are orthogonal to
$\pivec$. The same statements apply to $\symmR $ because it has the same unique steady-state distribution.

For convenience, we define the matrix
\begin{align}
\MM_{ji}&:=W_{ji}\sqrt{\pi_{i}/\pi_{j}} \,.
\label{eq:appCdef}
\end{align}
We write the Hermitian and anti-Hermitian parts of $\MM$ as
\begin{align}
\AA :=\frac{1}{2}(\MM+\MM^{T})\qquad \BB :=\frac{1}{2}(\MM-\MM^{T}).
\label{eq:appABdef}
\end{align}
Introducing the diagonal matrix $D_{ji}=\delta_{ji}\sqrt{\pi_{i}}$,
we can express $\MM$ and $\AA$ as similarity transformations of $\RR$ and $\symmR$ respectively, $\MM=D^{-1} \RR D$ and $\AA=D^{-1}\symmR D$. It then follows that they share the same spectra~\citep[Cor.~1.3.4]{hornMatrixAnalysis1990},
\begin{align}
\eigvec^{\MM}=\eigvec^{\RR}\qquad\text{and}\qquad\eigvec^{\AA}=\eigvec^{\symmR }.
\label{eq:appeveq}
\end{align}

Note that $\AA$ is Hermitian, thus $\eigvec^{\AA}$ and $\eigvec^{\symmR}$ are real-valued. In addition, the second eigenvalue of $\AA$ 
obeys
the variational principle~\citep[Sec.~4.2]{hornMatrixAnalysis1990} 
\begin{equation}
\lambda_{2}^\AA=\max_{\vvec \in \mathbb{C}^n:\;\vvec\perp\sqrt{\pivec},\Vert\vvec\Vert=1}{\vvec}^{\dagger}\AA{\vvec}\,,\label{eq:barA}
\end{equation}
where $\sqrt{\pivec}:=(\sqrt{\pi_{1}},\dots,\sqrt{\pi_{n}})^{T}$ and the notation $\boldsymbol{a}\perp \boldsymbol{b}$  indicates orthogonality, $\sum_i a_i b_i = 0$. 

%\section{\MakeTextUppercase{Spectral gap results}}

\subsection{Derivation of first main result~\eqref{eq:boundFE}}
\label{app:mainresult}

We first plug the definitions of $\AA$ and $\BB$ from \eqref{eq:appCdef} and \eqref{eq:appABdef} into the definition of steady-state EPR, which appears as \eqref{eq:epr} in the main text. After a bit of rearranging, this gives % 
\begin{align}
\eprR 
 & =\sum_{i\neq j}2\sqrt{\pi_{i}\pi_{j}}|\BB_{ji}|\atanh\frac{|\BB_{ji}|}{\AA_{ji}}\nonumber \\
 & \ge\frac{2}{\constZ}\sum_{i\neq j}\AA_{ji}|\BB_{ji}|\atanh\frac{|\BB_{ji}|}{\AA_{ji}}\nonumber \\
 &\geq\frac{2}{\constZ}\Big(\sum_{i\neq j}\AA_{ji}|\BB_{ji}|\Big)\atanh\frac{\sum_{i\neq j}\BB_{ji}^{2}}{\sum_{i\neq j}\AA_{ji}|\BB_{ji}|}.\label{eq:bb4}
\end{align}
The second line uses $\sqrt{\pi_{i}\pi_{j}}\ge \AA_{ji}/\constZ$,
where $\constZ$ is defined in \eqref{eq:cdef}. The third line applies Jensen's
inequality to the convex function $\atanh x$ (for $x\geq0)$ with
weights $\AA_{ji}|\BB_{ji}|/(\sum_{i\neq j}\AA_{ji}|\BB_{ji}|)$. 
Applying the Cauchy--Schwarz inequality then gives % 
\begin{align}
\sum_{i\neq j}\AA_{ji}|\BB_{ji}|\le\sqrt{\sum_{i\neq j}\AA_{ji}^{2}}\sqrt{\sum_{i\neq j}\BB_{ji}^{2}}% 
=\constB\,\Vert \BB\Vert_{F},\label{eq:bb5}
\end{align}
where we used $\constB$ from \eqref{eq:normdef} and the Frobenious norm $\Vert \cdot \Vert_{F}$. 

We now combine \eqref{eq:bb4} and \eqref{eq:bb5}, while using that $x\atanh(y/x)$ is decreasing in $x$ for $x,y\geq0$. 
After  a bit of rearranging, this gives
\begin{align}
\frac{\eprR}{2\constZ}\geq\left(\frac{\constB}{\constZ}\right)^2\,\frac{\Vert \BB\Vert_{F}}{\constB} \atanh\frac{\Vert \BB\Vert_{F}}{\constB}.\label{eq:dd}
\end{align}

Next, we will use the following
%We prove this using a 
classic theorem by Kahan~\citep{kahanSpectraNearlyHermitian1975},
which we state here without proof and using our own notation.

\vspace{5pt}
\noindent \emph{Theorem. For Hermitian $\AA \in\mathbb{C}^{n\times n}$ and any
$\BB\in\mathbb{C}^{n\times n}$, 
\begin{align*}
 & \Vert\eigvecW \AA-\rre\eigvecW{\AA+\BB}\Vert\le\\
 & \qquad\Big\Vert\frac{\BB+\BB^{\dagger}}{2}\Big\Vert_{F}+\sqrt{\Big\Vert\frac{\BB-\BB^{\dagger}}{2}\Big\Vert_{F}^{2}-\Vert\imag\eigvecW{\AA+\BB}\Vert^{2}}.
\end{align*}
} 

Observe that $\AA+\BB=\MM$, which follows from our definitions~\eqref{eq:appABdef}. 
Moreover, $\BB$ is anti-Hermitian, so $\Vert(\BB+\BB^{\dagger})/2\Vert_{F}=0$ and $\Vert(\BB-\BB^{\dagger})/2\Vert_{F}=\Vert \BB\Vert_{F}$.
Combining with the theorem above and rearranging gives 
\begin{align}
\Vert \BB\Vert_{F}^{2} 
& \ge\Vert\eigvecW \AA-\rre\eigvecW{\MM}\Vert^{2}+\Vert\imag\eigvecW{\MM}\Vert^{2}\nonumber \\
 & =\Vert\eigvecW \AA-\eigvecW{\MM}\Vert^{2}=\big\Vert\eigvecW{\RR}-\eigvecW{\symmR}\big\Vert^2 .\label{eq:appEq2}
\end{align}
In the last line, we used that $\imag\eigvecW \AA=0$ since $\AA$
is Hermitian, and then used Eq.~\eqref{eq:appeveq}.

The result \eqref{eq:boundFE} follows by combining Eqs.~\eqref{eq:dd} and \eqref{eq:appEq2},
while using that $x \atanh x$ is decreasing in $x \ge 0$. 

\subsection{Derivation of the spectral gap inequality \eqref{eq:fffd}}

\label{app:secev}

% Next, 
Let $\uvec$ be the right eigenvector of $\MM$ corresponding to $\eigvecWi{\MM}2$, 
normalized so that $\Vert\uvec\Vert=1$. Using $\MM \uvec=\eigvecWi{\MM}2\uvec$
and $\uvec^\dagger \MM^T=(\eigvecWi{\MM}2)^{*}\uvec^{\dagger}$, we can express the real
part of the eigenvalue as 
\begin{align}
\rre\eigvecWi{\MM}2=%\frac{\eigvecWi{\MM}2+(\eigvecWi{\MM}2)^{*}}{2}=
\frac{1}{2}\uvec^{\dagger}(\MM+\MM^{T})\uvec={\uvec}^{\dagger} \AA \uvec.\label{eq:reLA}
\end{align}

Since $\MM=D^{-1} \RR D$ and $\onevec^T\RR=0$, 
$\MM$ has a left eigenvector $(\onevec^T D)^T = \sqrt{\pivec}$ with the corresponding
eigenvalue $\eigvecWi{\MM}1=0$. The vector $\uvec$ is orthogonal to this eigenvector,  
$\uvec\perp\sqrt{\pivec}$. Therefore, $\uvec$ satisfies the constraints of the maximization~\eqref{eq:barA}. Combining \eqref{eq:appeveq}, \eqref{eq:barA}, and \eqref{eq:reLA} then gives 
\begin{align}
\eigvecWi{\symmR}2=\eigvecWi \AA 2 
\geq {\uvec}^{\dagger}\AA\uvec=\rre\eigvecWi{\MM}2=\rre\eigvecWi{\RR}2.
\label{eq:appsgap}
\end{align}
Finally, to derive~\eqref{eq:fffd}, note that $\symmR$ and $\RR$ are rate
matrices, thus $\eigvecWi{\symmR}2\le0$ and $\rre\eigvecWi{\RR}2\le0$.

We emphasize that the inequality~\eqref{eq:appsgap} is known in the literature on irreversible Monte Carlo methods~\citep{ichikiViolationDetailedBalance2013,suwa2010markov,bierkens2016non,diaconis2000analysis,turitsyn2011irreversible,chen2013accelerating,takahashi2016conflict,sakai2016eigenvalue,kaiser2017acceleration,ghimenti2022accelerating}. The derivation above is provided for completeness.

\vspace{10pt}

\subsection{Derivation of spectral gap inequalities \eqref{eq:ng2}}
\label{app:ng2}

Using $\vert\lambda_{2}^{\symmR }\vert=\vert\lambda_{2}^{\AA}\vert=-\lambda_{2}^{\AA}$ and the variational principle~\eqref{eq:barA}, we write 
\begin{align*}
\vert\lambda_{2}^{\symmR }\vert
& =-\max_{\vvec:\,\vvec\perp\sqrt{\pivec},\Vert \vvec\Vert=1}\sum_{ij}v_{i}^{*}\AA_{ij}v_{j}\\
 & =-\max_{\vvec:\,\vvec\perp\sqrt{\pivec},\Vert \vvec\Vert=1}\sum_{ij}\frac{v_{i}^{*}v_{j}}{\sqrt{\pi_{i}\pi_{j}}}\symmR _{ij}\pi_{j},
 \end{align*}
 where we used that $ \AA_{ij}=\symmR_{ij}\sqrt{\pi_{j}/\pi_{i}}$, as follows from our definitions \eqref{eq:wbardef}, \eqref{eq:appCdef}, and \eqref{eq:appABdef}. 

 Next, we introduce the variables $$\phi_{i}:=v_{i}/\sqrt{\pi_{i}}$$
and the norm $$\Vert\phivec\Vert_{\pivec}=\sqrt{\sum_{i}\vert\phi_{i}\vert^{2}\pi_{i}}\,.$$ 
This allows us to rewrite
\begin{align*}
\vert\lambda_{2}^{\symmR }\vert & =-\max_{\phivec:\,\phivec\perp\pivec,\Vert\phivec\Vert_{\pivec}=1}\sum_{ij}\phi_{i}^{*}\phi_{j}\symmR _{ij}\pi_{j}\\
 & =\frac{1}{2}\min_{\phivec:\,\phivec\perp\pivec,\Vert\phivec\Vert_{\pivec}=1}\sum_{i\neq j}|\phi_{i}-\phi_{j}|^{2}\symmR _{ij}\pi_{j},
\end{align*}
%where the third equality follows by substituting $\phi_{i}=v_{i}/\sqrt{\pi_{i}}$
%and defining $\Vert\phivec\Vert_{\pivec}=\sum_{i}\vert\phi_{i}\vert^{2}\pi_{i}$,
%and the last equality 
where the second line follows from $\symmR _{ii}=-\sum_{j:\,j\neq i}\symmR _{ji}$
and the detailed balance condition $\symmR _{ij}\pi_{j}=\symmR _{ji}\pi_{i}$. Using 
\[
\symmR _{ij}\pi_{j}=\frac{\RR_{ij}\pi_{j}+ \RR_{ji}\pi_{i}}{2\pi_{i}\pi_{j}}\pi_{i}\pi_{j}\leq\kappa G_{ij}\pi_{j}\leq\kappa\pi_{i}\pi_{j},
\]
we can bound $\vert\lambda_{2}^{\symmR }\vert$ as
\begin{align*}
\vert\lambda_{2}^{\symmR }\vert & \leq\frac{\kappa}{2}\min_{\phivec:\,\phivec\perp\pivec,\Vert\phivec\Vert_{\pivec}=1}\sum_{i\neq j}|\phi_{i}-\phi_{j}|^{2}G_{ij}\pi_{j}=\kappa\vert\lambda_{2}^{G}\vert\\
 & \leq\frac{\kappa}{2}\min_{\phivec:\,\phivec\perp\pivec,\Vert\phivec\Vert_{\pivec}=1}\sum_{i\neq j}|\phi_{i}-\phi_{j}|^{2}\pi_{i}\pi_{j}=\kappa,
\end{align*}
where the last equality follows from $\frac{1}{2}\sum_{ij}|\phi_{i}-\phi_{j}|^{2}\pi_{i}\pi_{j}=1$
for any $\phivec$ that satisfies $\sum_{i}\phi_{i}\pi_{i}=0$ and $\sum_{i}\vert\phi_{i}\vert^{2}\pi_{i}=1$.
This recovers the inequalities~\eqref{eq:ng2}.

\bibliographystyle{IEEEtran}
\bibliography{writeup}

\vfill

\clearpage
\end{document}